\newif\ifpnas
\pnasfalse

\ifpnas
\documentclass[9pt,twocolumn,twoside]{pnas-new}
    \usepackage{dsfont}
    
\usepackage{verbatim}
\usepackage{amsmath}
\usepackage{comment}
\usepackage{amsfonts}
\usepackage{amssymb}
\usepackage{numprint}
\usepackage{bm,bbm}
\usepackage{fancyhdr}
\usepackage{empheq}
\usepackage{color} 
\else
  \documentclass[twocolumn,reprint,floatfix]{revtex4-1}
  \usepackage[utf8]{inputenc}
  \usepackage{newtxtext}
  \RequirePackage[dvipsnames,usenames]{xcolor}
  \usepackage{hyperref}

  \usepackage{color} 
  \usepackage{amsmath} 
  \usepackage{graphicx}

\usepackage{cancel}
\usepackage{url}
\usepackage{xcolor}
\usepackage{bm,bbm}
\usepackage{fancyhdr}
\usepackage{empheq}

\usepackage{graphicx}
\usepackage{amsmath}
\usepackage{amsfonts}
\usepackage{amssymb}
\usepackage{numprint}

\fi

\ifpnas
    \templatetype{pnasresearcharticle} 
\else
  \makeatother
 \begin{document}
\fi

\title{Macroscopic Magnetic Monopoles in a 3D-Printed Mechano-Magnet}
\ifpnas


\author[a]{H.\ A.\  Teixeira}
\author[a]{M.\ F.\ Bernardo}
\author[b,c]{M. D. Saccone}
\author[b]{F.\  Caravelli}
\author[b]{C.\  Nisoli}
\author[a]{C.\ I.\ L.\ Araujo}

\affil[a]{ Laboratory of Spintronics and nanomagnetism (LabSpiN), Departamento de F\'{i}sica,
Universidade Federal de Vi\c cosa, Vi\c cosa, 36570-900, Minas Gerais, Brazil }

\affil[b]{Center for Nonlinear Studies,
Los Alamos National Laboratory, Los Alamos, New Mexico 87545, USA}
 \affil[c]{ Theoretical Division (T4),
Los Alamos National Laboratory, Los Alamos, New Mexico 87545, USA}


\leadauthor{C. de Araujo}

\significancestatement{ Spin ice emergent magnetic monopoles appear at the atomic scale in rare earth pyrochlores or at the nano-scale in  artificial spin ices systems. We demonstrate that the notion of magnetic monopoles can be transported at the macroscopic scale. This is the first experiment demostrating macroscopic  emergent magnetic monopoles using 3D printed materials, which is readily accessible to experimentation without a sophisticated experimental apparatus.}

\authorcontributions{C.\ I.\ L.\ Araujo conceived of the project, built the system and supervised the experiments and the entire project. H.\ A.\  Teixeira and M.\ F.\ Bernardo performed the experiments and analyzed the data, under supervision of C.I.L. Araujo. M. Saccone and F. Caravelli developed a corroborating simulation code. C. Nisoli provided analysis and conceptualization of the data and worked with C.\ I.\ L.\ Araujo at the first draft. All authors contributed to the final manuscript. }
\authordeclaration{No conflict of interest.}
\correspondingauthor{\textsuperscript{2}To whom correspondence should be addressed. E-mail: caravelli@lanl.gov}


\else

\author{H.\ A.\  Teixeira}
\affiliation{ Laboratory of Spintronics and nanomagnetism (LabSpiN), Departamento de F\'{i}sica,
Universidade Federal de Vi\c cosa, Vi\c cosa, 36570-900, Minas Gerais, Brazil }
\author{M.\ F.\ Bernardo}
\affiliation{ Laboratory of Spintronics and nanomagnetism (LabSpiN), Departamento de F\'{i}sica,
Universidade Federal de Vi\c cosa, Vi\c cosa, 36570-900, Minas Gerais, Brazil }
\author{M. D. Saccone}
\email{msaccone@lanl.gov} \affiliation{ Theoretical Division (T4) and Center for Nonlinear Studies,
Los Alamos National Laboratory, Los Alamos, New Mexico 87545, USA}
\author{F.\  Caravelli}
\email{caravelli@lanl.gov} \affiliation{ Theoretical Division (T4),
Los Alamos National Laboratory, Los Alamos, New Mexico 87545, USA}
\author{C.\  Nisoli}
\email{cristiano@lanl.gov} \affiliation{ Theoretical Division (T4),
Los Alamos National Laboratory, Los Alamos, New Mexico 87545, USA}
\author{C.\ I.\ L.\ Araujo}
\email{dearaujo@ufv.br} 
\affiliation{Laboratory of Spintronics and nanomagnetism (LabSpiN), Departamento de F\'{i}sica,
Universidade Federal de Vi\c cosa, Vi\c cosa, 36570-900, Minas Gerais, Brazil }

\fi

\begin{abstract}
The notion of magnetic monopoles has puzzled physicists since the introduction of Maxwell's Equations and famously Dirac  had hypothesized them  in the context of quantum mechanics. While they have proved experimentally elusive as elementary particles, the concept has come  to describe excitations or topological defects in various material systems, from liquid crystals, to Hall systems, skyrmion lattices, and Bose-Einstein condensate. Perhaps the most versatile manifestation of magnetic monopoles as quasiparticles in matter has been in so-called spin ice materials. There, they represent violations of the ice rule, carry a magnetic charge, and can move freely unbound.  We have built a mechano-magnet realized via 3D-printing, that consists of mechanical rotors on which macroscopic magnets can pivot. By controlling the relative height of the rotors we can achieve different regimes for magnetic monopoles, including the free monopole state. We then explore their driven dynamics under field. In the future, integration of our proof of principle in an elastic matrix can lead to novel macroscopic mechano-magnetic materials, to explore unusual piezomagnetism and magnetostriction, with applications to actuators and soft-robotics. 
\end{abstract}

\ifpnas
  \dates{This manuscript was compiled on \today}
  \doi{\url{www.pnas.org/cgi/doi/10.1073/pnas.XXXXXXXXXX}}

  \begin{document}

  \maketitle
  \thispagestyle{firststyle}
  \ifthenelse{\boolean{shortarticle}}{\ifthenelse{\boolean{singlecolumn}}{\abscontentformatted}{\abscontent}}{}





\else
  \maketitle
  \section{Introduction}

\fi

While electrical dipoles can be broken into separable, opposite charges, breaking a magnetic dipole does not separate its north and south poles. It results  instead in two smaller dipoles. The possibility of elementary, quantum, magnetic monopoles has been explored theoretically by Dirac~\cite{dirac1931quantised}, but no such particle was ever found~\cite{preskill1984magnetic}. 
The concept, however, has proved useful in describing emergent quasi-particles in liquid crystals~\cite{chuang1991cosmology}, skyrmion lattices~\cite{milde2013unwinding}, Bose-Einstein condensates~\cite{ray2014observation}, and in the context of the anomalous Hall effect~\cite{fang2003anomalous}. 
The notion of a magnetic monopole is perhaps the closest to the classical intuition in a special class of frustrated binary magnets called spin ices, either crystal grown~\cite{Ramirez1999,Bramwell2001,fennell2009magnetic,ryzhkin2005magnetic,Castelnovo2008,morris2009dirac,bramwell2009measurement,bovo2013brownian} or artificially fabricated at the nanoscale, the so-called ``artificial spin ices"~\cite{Wang2006,skjaervo2019advances,mol2009magnetic,Mengotti2010,ladak2011direct,Nascimento2012,perrin2016extensive,farhan2019emergent,goryca2021field,modifiers,caravellimod}. These magnetic materials are made of interacting moments that can be described as binary variables arranged along the edges of a lattice. The moments are atomic in rare earth pyrochlores, and have sizes of tens or hundreds of nanometers in nanofabricated artificial spin ices. 

\begin{figure*}[hbt!]
\center
\includegraphics[width=12cm]{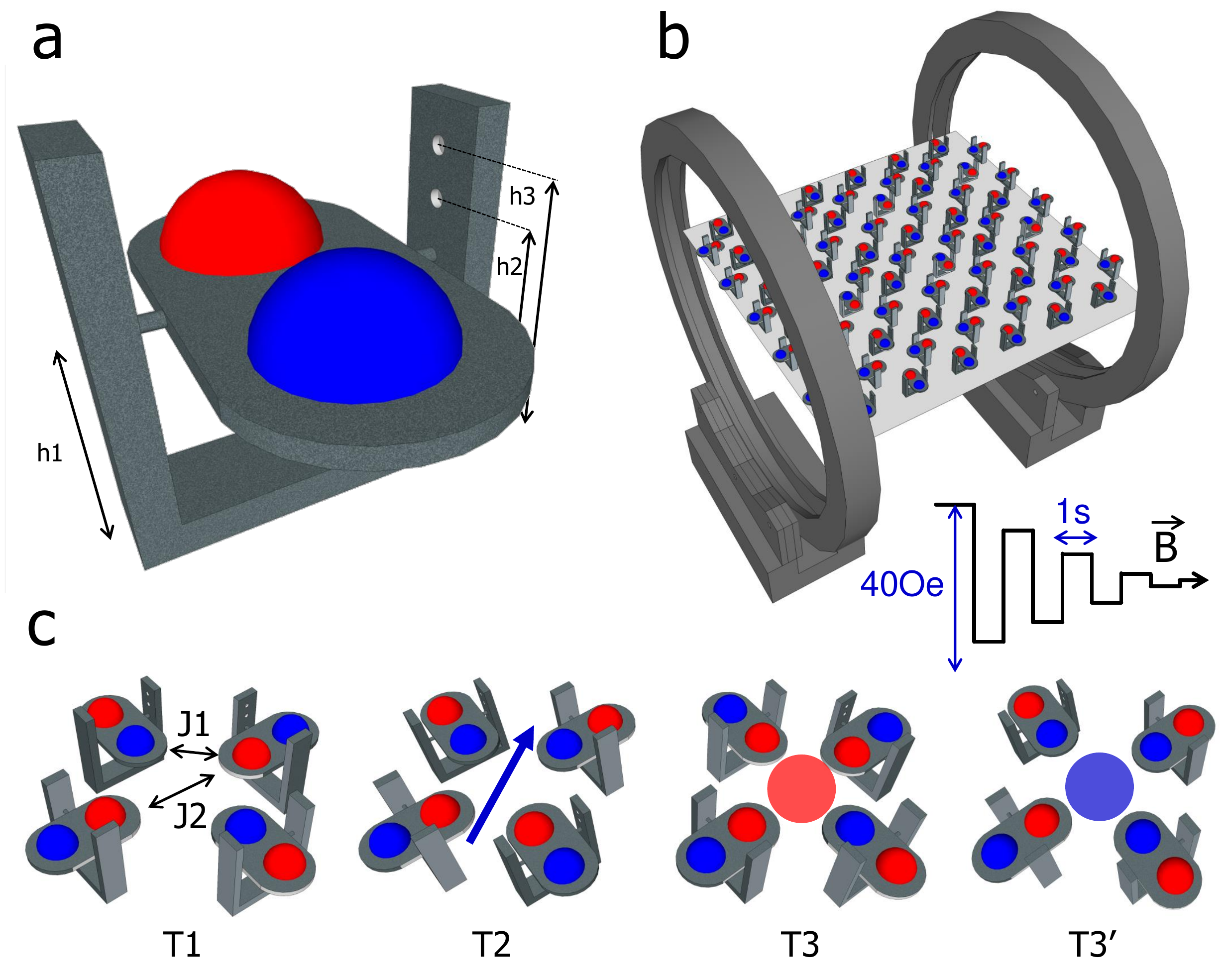}
\caption{{\bf The system}. a)  Design of the 3D-printed rotor, conceived to act as a dipole in a macroscopic artificial spin ice system. Each dipole is composed of a main axis of 15mm length, able to rotate in the polar direction, and two neodymium magnets of 5mm diameter.  The height of the axis can be changed in three offsets, in multiples of $5mm$. b) 84 rotors arrayed  along the geometry of a square spin ice, fit at the center of an Helmholtz coil of $20 cm$ radius. The inset shows the evolution of magnetic field in function of time in the demagnetization protocol. 
c) The observed vertex configurations. $T_1, T_2$ obey the ice rule (2-in/2-out), of which $T_2$ is magnetized along the blue line.  The $T_3$ configuration violates the ice rule and carries a charge $\pm2$.}
\label{fig:schematics}
\end{figure*}

We show here for the first time that magnetic monopoles  can be realized at the macroscale, in a mechano-magnet  made  of dipoles pivoting on mechanical rotors, shown in Fig.~\ref{fig:schematics}, a method pioneered before for kagome spin ice~\cite{mellado2012macroscopic} and non-degenerate square ice~\cite{gonccalves2020naked}. We develop the approach for a square lattice, and  incorporate the relative offset~\cite{Moller2006} known to achieve degeneracy and thus free monopoles~\cite{chern2014realizing,perrin2016extensive,farhan2019emergent}. 

\begin{figure}[hbt!]
\center
\includegraphics[width=8 cm]{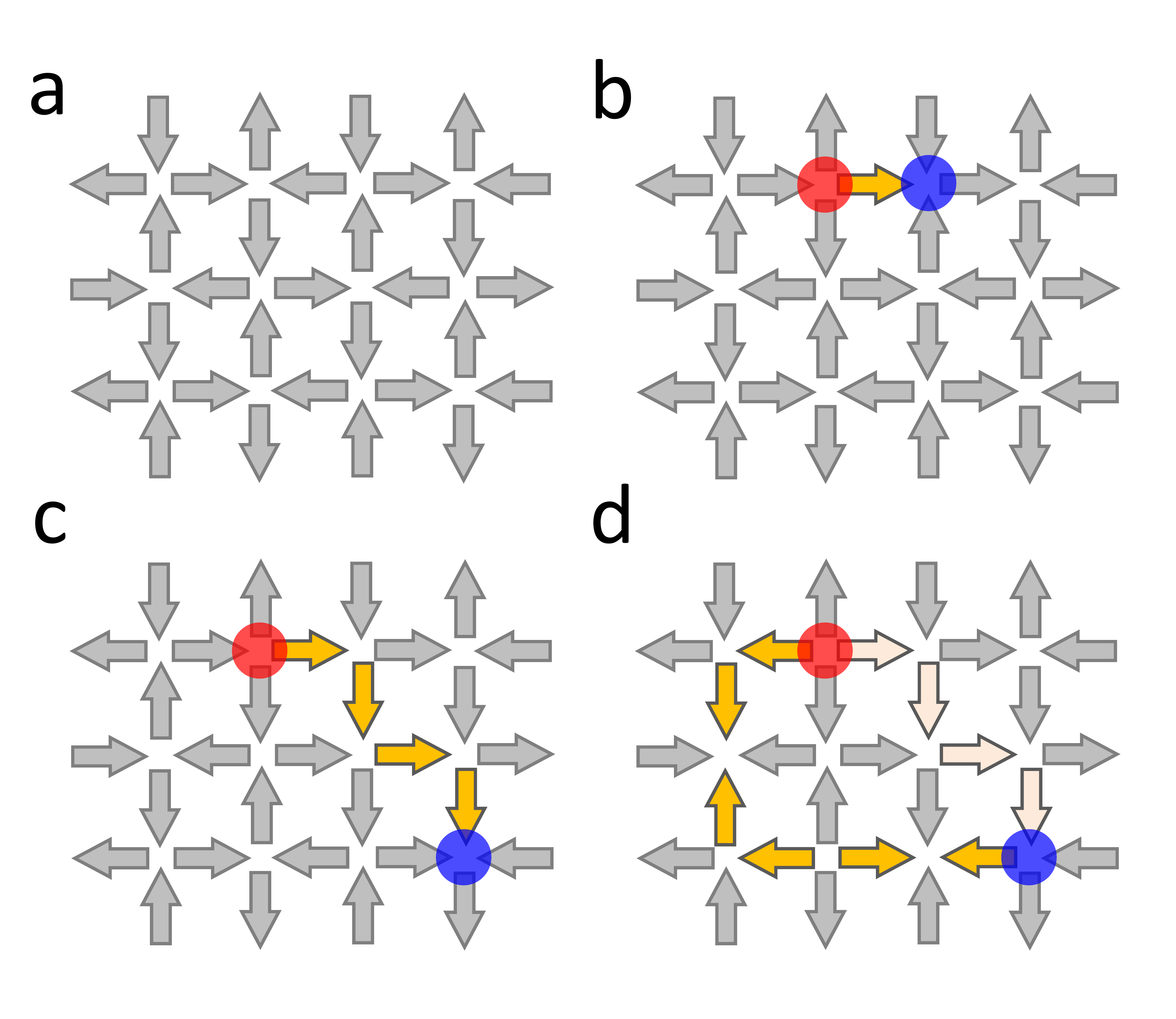}
\caption{{\bf Schematics of magnetic monopole excitations as violations of the ice rule}.  a) An example of ice rule configuration for Square Spin Ice: moments are represented by arrows; in each vertex, two moments point in, two point out. The number of such configuration scales exponentially with the number of moments, leading to a finite Pauling entropy per moment. b) Flipping one moment ({yellow}) in an ice rule configuration breaks the ice rule, and creates a monopole pair: a $q=+2$ monopole corresponding to three moments pointing in and one pointing out ({blue circle}), and a $q=-2$ monopole corresponding to three moments pointing out and one pointing in (\textcolor{black}{red circle}). c) Further flips of moments can separate the monopoles, leaving behind a so-called "Dirac String" (yellow spins), which is however not univocally determined. d) A different possible Dirac string is marked in yellow, the previous one in light {yellow}. Note that flipping any entire Dirac string annihilates the monopole pair.}
\label{fig:SQI}
\end{figure}

To understand our system, consider the model  {\it degenerate square spin ice} of Fig.~\ref{fig:SQI}. The low energy state is degenerate, and corresponds to  disordered configurations of moments that obey the so-called ice-rule~\cite{bernal1933theory,pauling1935structure} in which two moments point in, two point out of each vertex. 
If we define as charge of a vertex the number of moments pointing in minus the number of moments pointing out, then ice-rule vertices have zero magnetic charge. But flipping one moment creates a pair of opposite $\pm2$ charges, or magnetic monopoles~\cite{ryzhkin2005magnetic,Castelnovo2008}, that break the ice rule. Further flips can separate them without more violations of the ice rule and thus without an energy cost~\cite{Castelnovo2008}. Note  that these monopoles are in fact north and south poles of a long floppy dipole of sort, a magnetized line of moments (e.g. in grey in Fig.~\ref{fig:SQI}) often called ``Dirac string". However, because the manifold is disordered, other  strings of head-to-toe moments starting in a monopole and ending in another could be equally chosen as Dirac strings (Fig.~~\ref{fig:SQI}d), which are therefore not uniquely defined. This leads to the interpretation of the low energy manifold in terms of free monopole excitations of an ice rule obeying spin vacuum.

Analogous magnetic monopoles had been proposed in pyrochlore spin ices where they have also motivated a search for their current, or magnetricity~\cite{bramwell2009measurement}. They were recently revealed in square artificial spin ice ice at the nanoscale~\cite{perrin2016extensive,farhan2019emergent}.  Early realizations~\cite{Wang2006} possessed an ordered, antiferromagnetic ground state because moments converging perpendicularly in a vertex interact more strongly than those converging collinearly, thus lifting the degeneracy of the ice rule. However, degeneracy can be regained by offsetting horizontal and vertical magnets in the $z$ direction~\cite{Moller2006,chern2014realizing,perrin2016extensive,farhan2019emergent}. 

There are obviously enormous  differences in system, geometry, dynamics, scale, and interaction, between atomic magnetic moments in pyrochlore spin ice,  superparamagnetic nanoislands at thermal equilibrium, and macroscopic magnets pivoting on a rotor. Yet, we  show here that  their behavior can be conceptualized in similar ways. We implement the square ice geometry with magnetic rotors and an offset perpendicular to the plane. This allows us to explore in a macroscopic system the three different structural phases of square ice--the antiferromagnetic state, the degenerate state, and the line state--and the driven dynamics of their monopoles.


\begin{figure*}[ht!]
\center
\includegraphics[width=15cm]{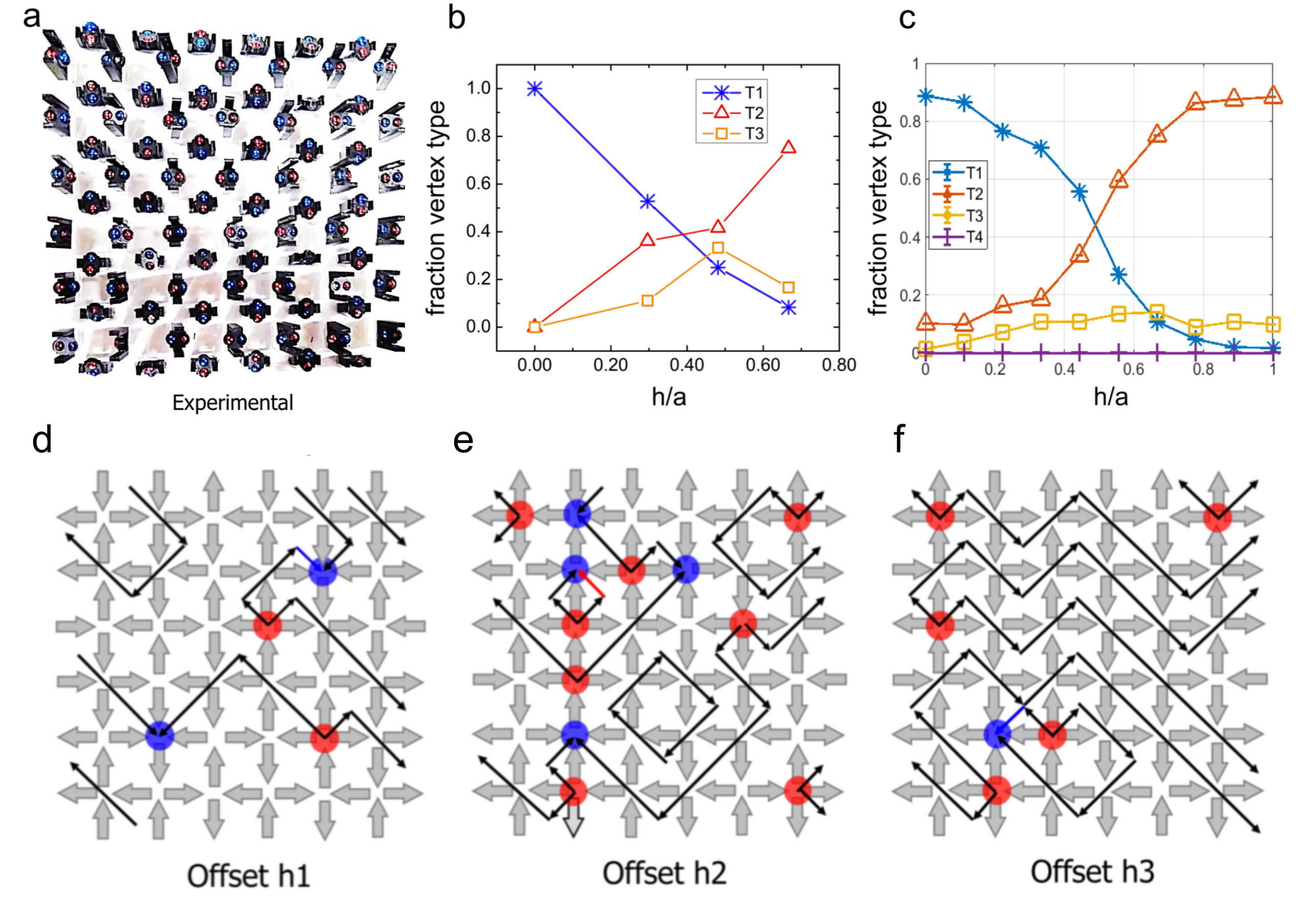}
\caption{{\bf AC Annealing of the mechano-magnetic spin ice.} a) Snapshot of experimental setup. c) Vertex configurations observed after the AC demagnetization protocol. b) Fractions of vertex types obtained after the demagnetization protocol as a function of the offset show $T_1$ vertices being promoted at low values of $h/a$,  $T_2$ vertices being promoted at  high values of $h/a$. c) Fractions of vertex types obtained via a mechanical dynamical simulation of a much larger system agree qualitatively well with the experimental results. d-f) Real space configurations after AC demagnetization for offsets $h_1=5mm$ (d),  $h_2=10mm$ (e) and $h_3=15mm$ (f). }
\label{fig:field_lines}
\end{figure*}
\ifpnas
\section*{System}
\else
\section{System}
\fi
 We have utilized a 3D printer to fabricate rotors equipped with neodymium spherical magnets of 5mm diameter   that are allowed to spin in the polar angle direction, as depicted in Fig.~\ref{fig:schematics}a.  The rotors are arranged pointing along the edges of a square lattice of lattice constant $a=27mm$. 
 
 The pivoting rotors can be set at four different heights, differing by $5 mm$:  $h_0=0, h_1=5mm$, $h_2=10mm$, and $h_3=15mm$. The rotors converging in a vertex form four possible configurations, shown in Fig.~\ref{fig:schematics}c, and called $T_1, T_2,T_3,$~\cite{Wang2006}. $T_1, T_2$ vertices obey the ice rule, but  $T_1$ vertices have zero net magnetic moment. $T_3$ vertices corresponds to violations of the ice rule in the form of $q=\pm3 \mp1=\pm2$ monopoles. A fourth set of configurations, $T_4$ vertices, or monopoles of charge $q=\pm 4$,  correspond to all spin pointing in or all out, are highly energetic and  never observed. 
 
We follow the general recipe of Ref.~\cite{Moller2006} and offset  the vertically aligned moments with respect to the horizontally aligned ones, to achieve different interaction strengths among perpendicular and collateral rotors ($J_1, J_2$ in Fig.~\ref{fig:schematics}c).  For an offset of zero, $T_1$ vertex configurations have the lowest energy, and thus the lowest energy of the array corresponds to an antiferromagnetic tessellation of $T_1$ vertices. For very high offset, the magnetized $T_2$ vertices are expected to be lowest in energy, and the collective lowest energy state is expected to be a so-called {\it line state}, as it corresponds to magnetization lines flowing through the system~\cite{perrin2016extensive,king2021qubit,nisoli2021gauge}. 

At some intermediate offset, by reasoning within a simplified nearest neighbor model for simple binary moments, one expects that the ground state of the system corresponds to the degenerate six-vertex model, and that its excitations correspond to free monopoles~\cite{Moller2006}. While our mechanical system is more complex than such model of binary spins,  our results confirm the same picture, as we shall show.

\ifpnas
\section*{Annealing via AC Demagnetization}
\else
\section{Annealing via AC Demagnetization}
\fi
For each of the four offsets, we prepared the system via an AC demagnetization protocol~\cite{ke2008energy,rodrigues2013efficient}. The external magnetic field generated by a Helmholtz coil with radius $R=20cm$ decreases in oscillatory  steps of amplitude $B_x=$ {5} Oe starting from a field of 40 Oe. Results obtained after the demagnetization processes are presented in Fig.~\ref{fig:field_lines}. In all cases ice-rule vertices are promoted, while monopoles are suppressed.

\begin{figure*}[t!]
\centering
\includegraphics[width=17cm]{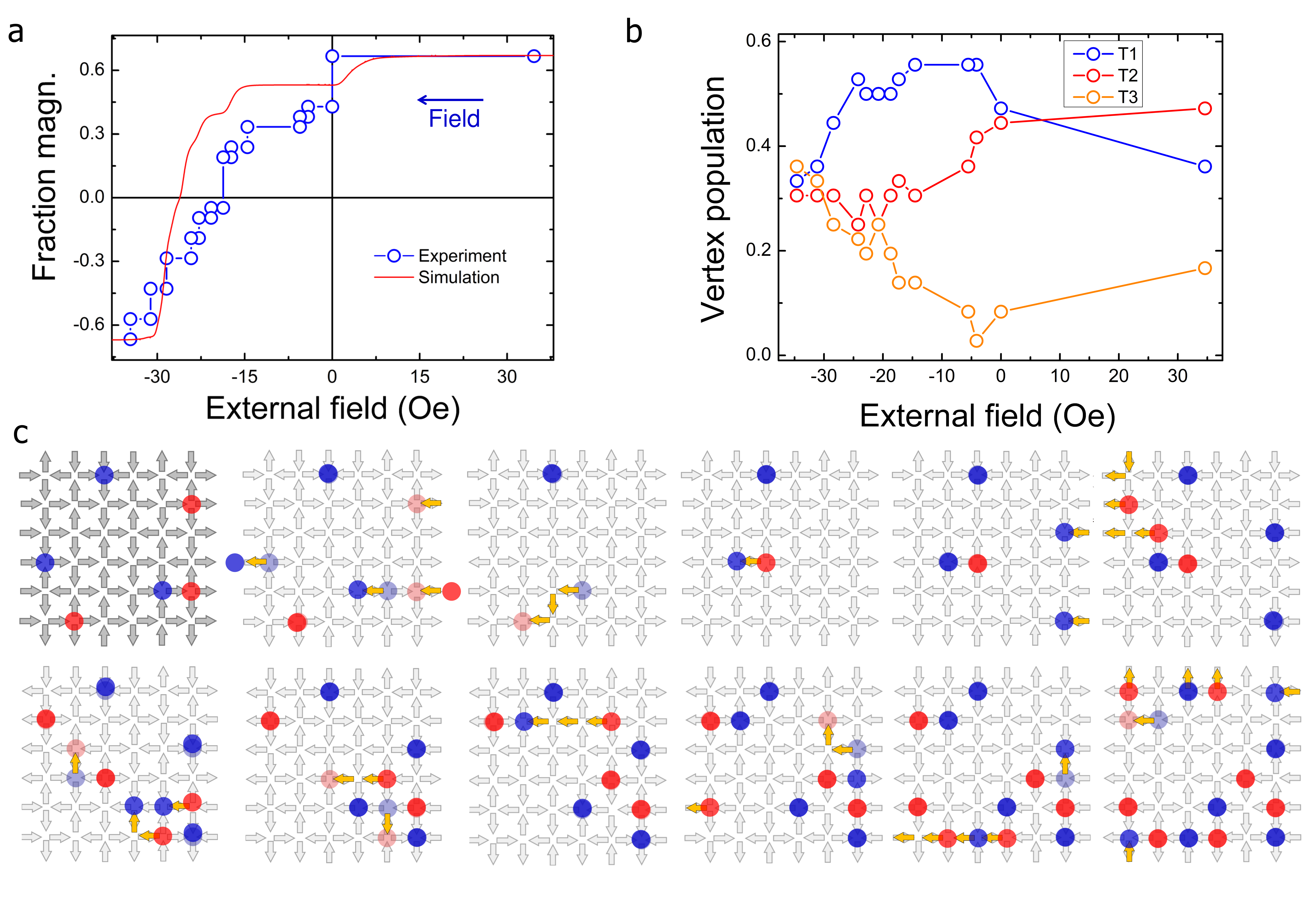}
\caption{{\bf Driving Macroscopic monopoles by Field Inversion in an anti-ferromagnetic spin ice (offset $h_1$).} a) Magnetic moment measured in units of rotors vs. applied field. b) Vertex statistics vs. applied field. Note the proliferation of $T_1$ antiferromagnetic vertices. c) Real-space snapshots of moment inversion. Yellow moments have flipped with respect to the previous frame.}
\label{FigHyst1}
\end{figure*}

\begin{figure*}[t!]
\centering
\includegraphics[width=16cm]{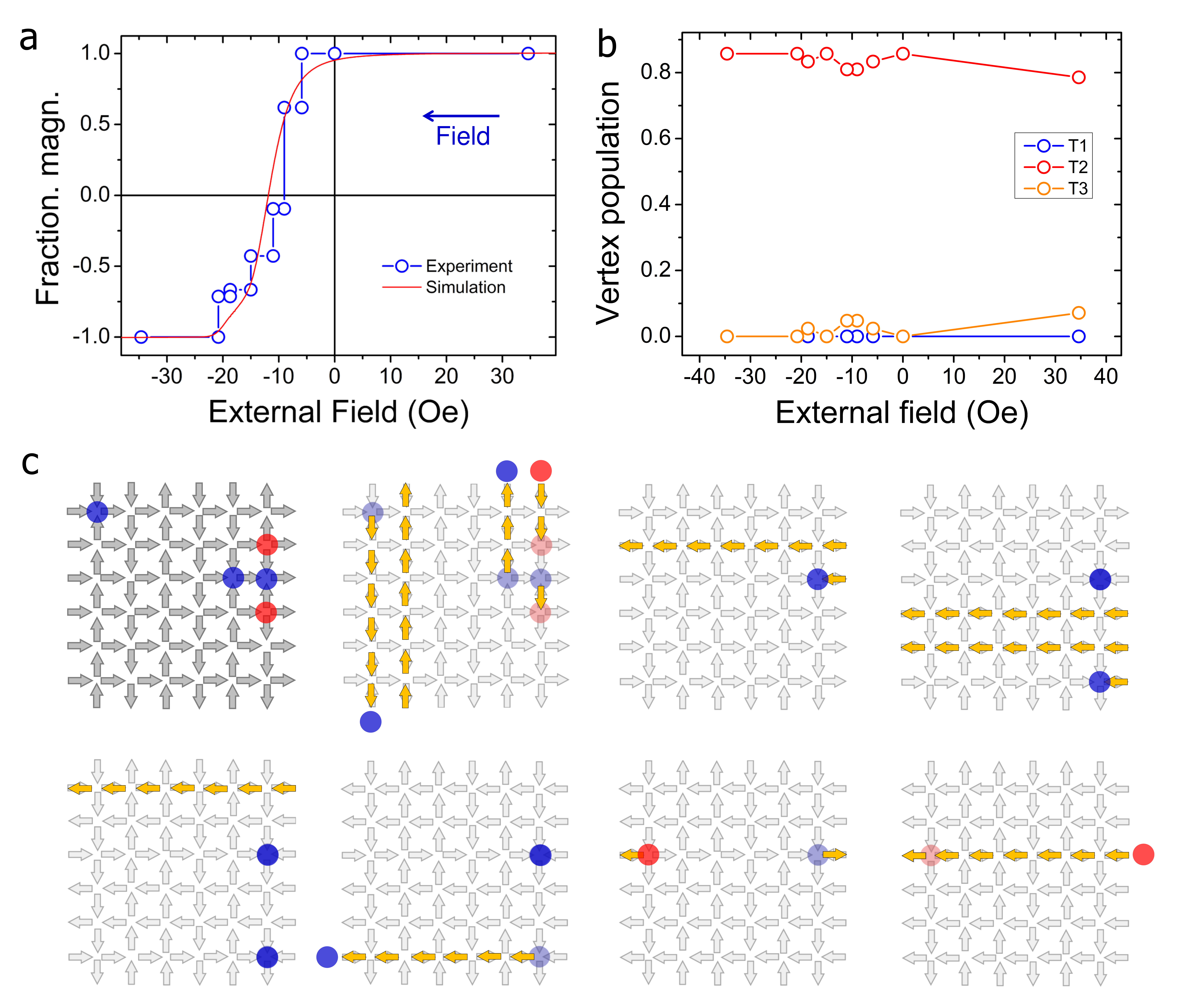}
\caption{{\bf Driving Macroscopic monopoles by Field Inversion in a line state spin ice (offset $h_3$).} a) Magnetic moment measured in units of rotors vs. applied field. b) Vertex statistics vs. applied field. Note the absence of $T_1$ configurations, as $T_2$s transform into monopoles ($T_3$)  and then into $T_2$ configuration of opposite magnetic moment. }
\label{FigHyst3}
\end{figure*}

For zero offset we obtain a completely antiferromagnetic configuration, the expected tessellation of $T_1$ vertices of zero net magnetic moment. At the offset $h_1$,  the energy difference between the $T_2, T_3$  and the $T_1$ vertex configurations is reduced, and  we obtain antiferromagnetic domains of $T_1$ vertices, as  shown in the schematics of  Fig.~\ref{fig:field_lines}d, in which  the rotors are represented by arrows pointing in the direction of their projection in the plane. Note that, unlike the binary variables of standard spin ices, here the rotors have a continuous degree of freedom: their pivoting angle. The domains are separated by domain walls,  which we draw according to the convention of Ref.~\cite{nisoli2020topological} by joining together the magnetic moment of $T_2$ vertices. These domains are also ``Faraday Lines''~\cite{nisoli2020topological} entering negative monopoles and exiting positive ones. The obtained configuration has a small net magnetization per moment  $Mx_{h1}=\textcolor{black}{0.095}$, defined  by assigning to each arrow a moment $\pm1$.

For the largest offset $h_3$ the lowest energy configuration corresponds to $T_2$ vertices. Indeed,  Fig.~\ref{fig:field_lines}b,d shows that  most vertices are in the magnetized $T_2$ configuration. The Faraday lines no longer describe domain walls, but a line state that covers almost the entire sample, which has now a total magnetic moment \textcolor{black}{$M_x=0.42$, $M_y=0$}, as also seen previously for nanomagnetic realizations~\cite{perrin2016extensive}. 

At some intermediate offset, the  $T_1$ and $T_2$ configurations are predicted to become degenerate. For ideal dipoles, this critical height is estimated at $h/a\simeq 0.419$, but  the finite length of the dipole lowers that ratio. In our case, from $h_c= a-l/\sqrt{2} $, using the length of the rotor $l=1.2mm$ and lattice constant $a=27 mm$ return the offset $h\simeq10 mm$.  corresponding to our $h_2$. 

The results of the annealing at offset $h_2$ is shown in Fig.~\ref{fig:field_lines}c. The sample is fairly demagnetized, with {$M_x=-0.047$, $M_y=0.143$}. Note that the ratio of relative occurrence among $T_2$ and $T_1$ vertices is $\simeq 1.66$, quite close to the ratio  among their multiplicities $2=4/2$ (there are four types of $T_2$ vertices,  two types of $T_1$ vertices) and similar to that observed in literature for nanomagnetic systems either demagnetized via AC demagnetization or in thermal equilibrium \cite{mol2010conditions,perrin2016extensive,farhan2019emergent}. Interestingly, the observed frequency of vertex configurations corresponds to that found in a {\it thermal state} for $T/J=1.1$, as it can be  verified via a  Metropolis Monte Carlo simulation of nearest neighbors coupling $J$. This is not completely surprising, as it has been shown before that AC demagnetization can lead to ensembles similar to the thermal ones~\cite{Nisoli2007,Nisoli2010,perrin2019quasidegenerate}. 

In Supplementary Materials, we show how to simulate our system within a driven mechanical model of rotors. Our numerical results agree well with the experimental findings, as shown in Fig.~\ref{fig:field_lines}c. 

Finally, we note that the geometric Gauss's law is fulfilled in all samples. If we call $\phi$ the flux of the magnetic moments inside the system, defined as number of boundary spins pointing in minus those pointing out, than $\phi$ must be equal the net charge in the system, or $\phi =Q_{\text{tot}}$. That number is zero for the offset of zero and $5mm$, the antiferromagnetic case, and $-8$ in the other two cases.

\begin{figure*}[t!]
\centering
\includegraphics[width=16cm]{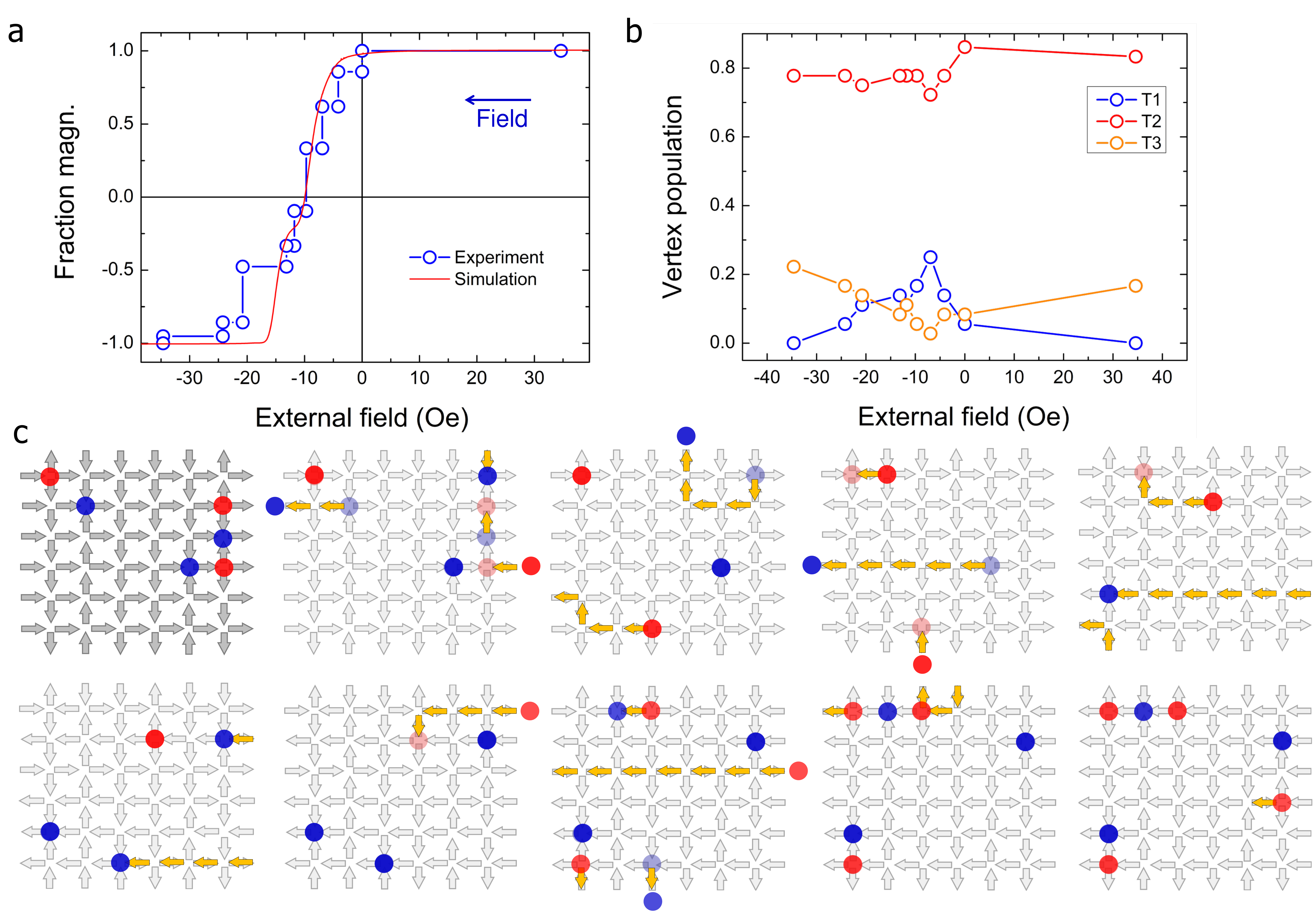}
\caption{{\bf Driving Macroscopic monopoles by Field Inversion in a degenerate spin ice state (offset $h_2$).} a) Magnetic moment measured in units of rotors vs. applied field. b) Vertex statistics vs. applied field. }
\label{FigHyst2}
\end{figure*}

\ifpnas
\section*{Monopoles Driven by Field}
\else
\section{Monopoles Driven by Field}
\fi
We explore monopole dynamics by driving them with an in-plane external magnetic field in the $[1,0]$ direction. We saturate the material under a field of $\sim 35$ Oe and then invert the field. When field is inverted linearly, the behavior of the system proceeds by short bursts of collective rearrangement of spins. Each time the system changes, we pause the field inversion and collect data at that field. Therefore, in the data shown in Fig.~\ref{FigHyst1}, Fig.~\ref{FigHyst3}, and Fig.~\ref{FigHyst2} the system remains unchanged as the field decreases between the various recorded data points, which are then shown as real space images.

In Fig.~\ref{FigHyst1} we report results for the offset $h_1$ (antiferromagnetic spin ice). For this case, the interaction is strong and a  field of 35 Oe cannot fully polarize it. Under inversion, at low negative field, $T_2$ vertices transform first into monopoles, and then into $T_1$ vertices, generating anti-ferromagnetic domains whose domain walls remain magnetized. Then as the negative field grows, the kinetics becomes largely dominated by the movement of domain walls. This effect can be seen in the curves for vertex configurations (Fig.~\ref{FigHyst1}b) and also in the magnetization curve (Fig.~\ref{FigHyst1}a). The latter is not as sharp as the other two cases (shown below), reflecting the complexity of the vertex transformations involved in the inversion of the overall moment. The $T_2$ vertices change their orientation via  elaborate set of  ``reactions''. Typically, for a vertex, the  reactions pathway is $T_2\to T_3 \to T_1 \to T_3 \to T_2$ and cannot always be completed. This leaves and abundance of magnetized monopoles and causes an asymmetry between the initial state polarized to the right and the final state polarized to the left. This asymmetry is absent in the other two cases.

In Fig.~\ref{FigHyst3} we show the case of the largest offset $h_3$, whose low energy manifold is a line state. At saturation horizontal  spins are all pointing in the same direction. Vertical spins form ferromagnetic rows of alternating direction, due to their antiferromagnetic interactions. Because of the large offset, and unlike in the $h_1$ case,  $T_1$ vertices have now higher energy than $T_2$. Thus  $T_1$ vertices do not appear as an intermediate step in the pathway of moment inversion. The moment inversion proceeds as $T_2 \to T_3 \to T_2$, that is via creation of monopoles ($T_3$) and their propagation through the system, which leaves in its wake a line of $T_2$ vertices properly aligned along the field. Sometimes, entire rows of moments flip. This can always be interpreted as a monopole ``entering'' the system from the right boundary and being pushed out at the left boundary. The monopoles, driven by the field, appear to propagate ballistically. Expectedly, the positive monopoles mostly move along the direction of the field, the negative ones in the opposite direction. The lack of the complex vertex-reactions of the previous case makes the magnetization profile  sharper (Fig.~\ref{FigHyst3}a). This allows for a reasonable, if approximated, determination of a coercive field at around 10~Oe. The statistics of vertex configurations (Fig.~\ref{FigHyst3}b) looks therefore considerably flatter than in the other cases, with a small maximum in monopole presence around the coercive field. At high negative field, the monopoles are eliminated by pushing them out of the boundaries.

The case of offset $h_2$, or degenerate square ice, is intermediate between the two others (Fig.~\ref{FigHyst3}). In the absence of a field, the $T_1$ and $T_2$ vertex configurations are energetically degenerate. However, in presence of a field $T_2$ vertices aligned along the field possess lower energy than the $T_1$ vertices, because of the Zeeman contribution from the field~\cite{gilbert2015direct,goryca2021field}. During reversal, the energetic advantage of the newly misaligned $T_2$ vertices is lost as they then oppose the external field. This might explain why the $T_1$ configurations reach a maximum at the coercive field (again at circa 10 Oe) where their energy is lower than most $T_2$ vertices. The initial slope in relative magnetization is large as the same ballistic monopole motion observed in the ferromagnetic system can quickly reverse lines of spins, even while $T_1$ vertices allow for more complex motion. There is a secondary leap in magnetization around 20 Oe as the population of $T_1$ vertices decrease while $T_3$ increase. This is the latter portion of the vertex reactions described for the antiferromagnetic spin ice, a $T_1 \to T_3 \to T_2$ process. The monopoles may move more freely and, with a lesser slope than the initial reversal, gradually convert away all the $T_1$ vertices while increasing the net magnetic moment.

Finally in the panels $a$ of Figs.~\ref{FigHyst1}-\ref{FigHyst2} we also show in red the simulated lines of magnetization after an annealing that is sinusoidal with a time- decreasing envelope. While there is no perfect agreement, as the real system is more complex to simulate mechanically, the simulations show similar qualitative differences among the three cases seen in experiments.

\ifpnas
 \section*{Discussion}
 \else
 \section{Discussion}
 \fi
We have provided proof of principle that subtle magnetism of degenerate frustrated  manifolds and magnetic monopoles can be realized at the macroscale, where macroscopic magnetic monopoles can be driven by external field. We have found that models and ideas developed at the atomic or nanoscopic scale are applicable to a much different macroscopic system. In particular, we have reproduced the three theoretically predicted regimes already realized at the nanoscale~\cite{Wang2006,perrin2016extensive,farhan2019emergent} and in the superconductive qubits of a quantum annealer~\cite{king2021qubit}, and similar monopole dynamics, but at the macroscale. 

Because the constrained disorder of the degenerate square ice (realized by us at offset $h_2$), this is another classical and macroscopic example of a seemingly topological phase~\cite{henley2010coulomb,castelnovo2007topological}, due to a kinetically constrained dynamics. One key difference is that, however, the real degrees of freedom of this system are continuous, and we did observe in simulations and experiments fast cascade transitions which reabsorb monopole pairs. Such fast cascades will be the subject of future experiments and numerical analysis. It is in fact typically the case that rapid cascades in system with a threshold dynamics leads to self-organized criticality and punctuated equilibria~\cite{soc}. Avalanches were seen in antiferromagnetic square ice \cite{bingham2021avalanches}.  Our system is an ideal platform to study monopole-related avalanches at the macroscale.

Our work also show another instance of apparently thermal ensembles achieved in an athermal system via external driving~\cite{d2003observing}. In the future it would be interesting to investigate relaxation procedures based on mechanical rather than magnetic drives, such as vibro-fluidization. Finally, in the context of the recent introduction of ideas from frustration and spin ice in the field of mechanical metamaterals~\cite{meeussen2019topological,pisanty2021putting,merrigan2021topologically}, our work can open new venues for macroscopic magneto-mechanical systems, of unusual magnetostriction or piezomagnetism.

\newcommand\ackcontent{
The authors thank CNPq, CAPES and FAPEMIG (Brazilian agencies) for
financial support.  The work of F. Caravelli, C. Nisoli and M. Saccone was carried out under the auspices of the NNSA of the U.S. DoE at LANL under Contract No. DE-AC52-06NA25396, financed via LDRD-ER grant PRD20190195. }

\newcommand\methods{The rotors presented in Figure 1 were developed by a conventional 3d printing process using Polylactic acid material. The main halter total length of 12mm is formed by two neodymium magnetic spheres with 5mm diameter each. This main part is connected to the external arms by a conic groove, allowing rotation with minimal friction. Three different distances among grooves were used to perform the z-axis offsets h=5,10 and 15mm discussed in the paper, with the halters aligned in y direction.
The rotors were disposed in a square lattice, at a distance of L=27mm, and a total length for the apparatus of 20cm. This size was used in order to ensure the uniformity of the external field along the experiment. The external field was applied by a commercial Helmholtz coil. 

To anneal the sample, we employed an oscillatory demagnetization protocol, starting with a current of 5A (the maximum allowed in the coil), corresponding to an external field of 35Oe; we then decreased the magnetic field by steps of 5Oe per second, and took a picture of the apparatus. The reversal processes were performed starting with system saturation, with a maximum field of 35Oe along the x-axis. We then mapped the population  density observed in each picture to a snapshot, as in the paper. A video of the experiment is available at a video repository. \footnote{Experimental video available at \href{https://www.youtube.com/watch?v=oxnfRsieK2A}{https://www.youtube.com/watch?v=oxnfRsieK2A}.}}

\ifpnas
\showmatmethods{\subsection*{Materials \& Methods}\methods} 

\acknow{\ackcontent}
\showacknow{} 

\else 
\section*{Acknowledgments}
\ackcontent

\fi

\bibliography{library2.bib}{}

\ifpnas
\section*{Supporting Information}
\else
\section*{Appendix}
\subsection{Experiment setup}
\methods
\fi

\ifpnas
\subsection*{Molecular dynamics simulations, annealing and monopoles movement}
\else
\subsection{Molecular dynamics simulations, annealing and monopoles movement}
\fi
We performed molecular dynamics simulations, by numerically implementing the force applied on each magnetic charge (approximating every magnetic rotor as a pair of charges) and projecting onto the axis of rotation.\footnote{We are happy to share the code upon request. Please write to caravelli@lanl.gov or msaccone@lanl.gov.} We performed a direct numerical integration of the equations via a Verlet method (e.g. the second order direct Euler discretization), with an integration time step $\delta t=10^{-2}$, in the absence of noise sources (zero temperature). The parameters used are those we inferred to best match the experimental results, as we describe below. The equations of motion for each rod were those derived in \cite{mellado2012macroscopic} with the addition of an external field coupling
\begin{eqnarray}
I\frac{d^2\alpha_{i}}{dt^2}&=&T_{i}-\eta \frac{d\alpha_{i}}{dt}+  \tau_i    \\ 
T_{i}&=&\left(\frac{a \mu_{0}}{4\pi}\right)\sum_{j\neq i}q_{i}q_{j}\hat{r}_{i}^{cm}\frac{\vec{r}_{ij}}{|\vec{r}_{ij}|^{3}}\\
\tau_i&=&\mu_B m(t) p_i^{\perp}\cdot (\vec{s}_i(\theta_i) \times \hat h);
\end{eqnarray}
In the equations above, the total torque is comprised of torques between nearest neighbor rotors, $T_{i}$, the kinetic friction coefficient is $\eta$, and external field torque $\tau_i$. $\vec{r}_{ij}$ is the vector joining the magnetic charges $q_{i}$ and $q_{j}$, while $\hat{r}_{i}^{cm}$ is unit vector pointing towards the charge $q_i$ from the center of mass of a rotor. 
The quantity $\tau_i$ is the torque force applied to the dipole, where $p_i^\perp$ is the vector associated with the dipole rotation axis, $\hat h$ is the direction of the external field, and $\vec s_i$ is the vector associated to the dipole; $m(t)$ is the time depedent magnitude of the field, and $\mu_B$ is the effective coupling between the dipole and external field.
We obtained a reasonable fit with the experimental results by choosing $\eta=1$, $a=1 $ (mm). We did not introduce static friction, which should explain the difference between the numerical and experimental results in the main text. $I$ was chosen so that in absence of forcing, the relaxation time, controlled by $\eta/I\approx 1$ is such that the dipoles relax within a timescale of $\approx 1$ second (a number of steps equal to $1/dt$), which is consistent with independent experimental observations. 
We have allowed for a staggered height of the monopoles on a checkerboard lattice (e.g. an island has height 0 and the neighbors have height $h$). Thus, the distance between two islands is given by $r_{ij}=\sqrt{d_{ij}^2+ (h_i-h_j)^2}$, where $d_{ij}$ is the distance on the plane.

\begin{figure}
    \centering
    \includegraphics[scale=0.19]{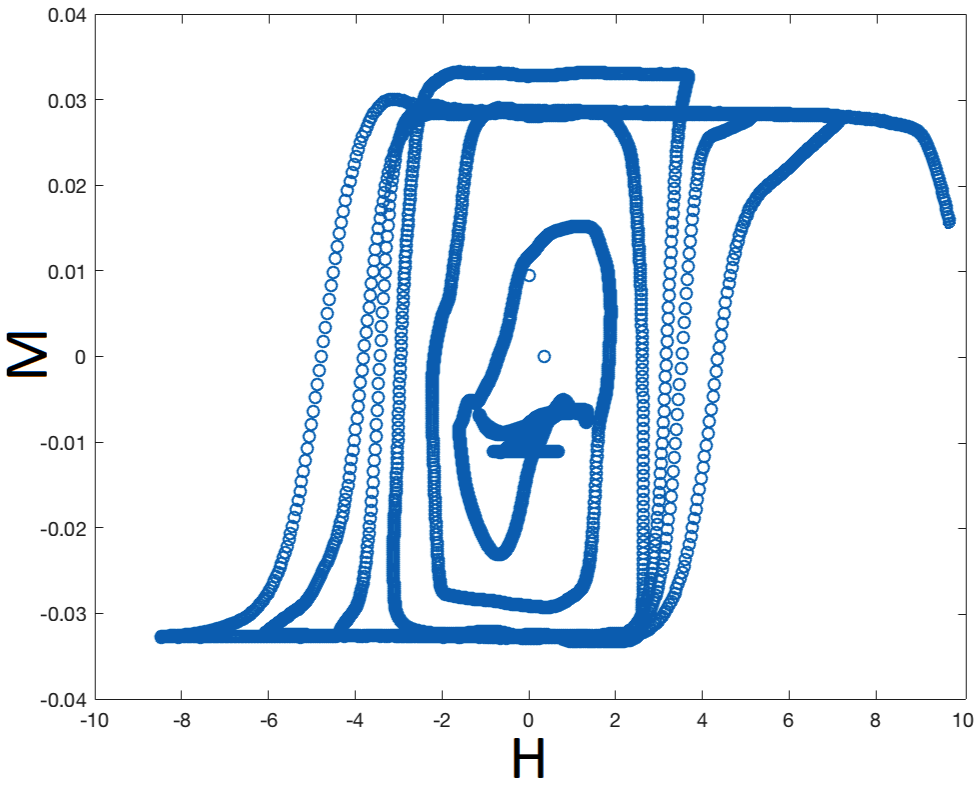}
    \caption{Numerical annealing of the macroscopic dipoles. The magnetization as a function of the external fields (in units of $\mu_B=1$). We see that the shape of the annealing resembles the experimental results. The magnetization is $M$ is the one parallel to the external field $\vec H=m(t) \hat H$.} 
    \label{fig:ann}
\end{figure}

First, we annealed the sample with a decaying, oscillating field applied diagonally as to affect both subsets of spins equally. We attempted to place the system in the ground state (at $h=0$ this corresponds to a Type I vertices). This can be seen in Fig. \ref{fig:ann}.

\begin{figure}
    \centering
    \includegraphics[scale=0.19]{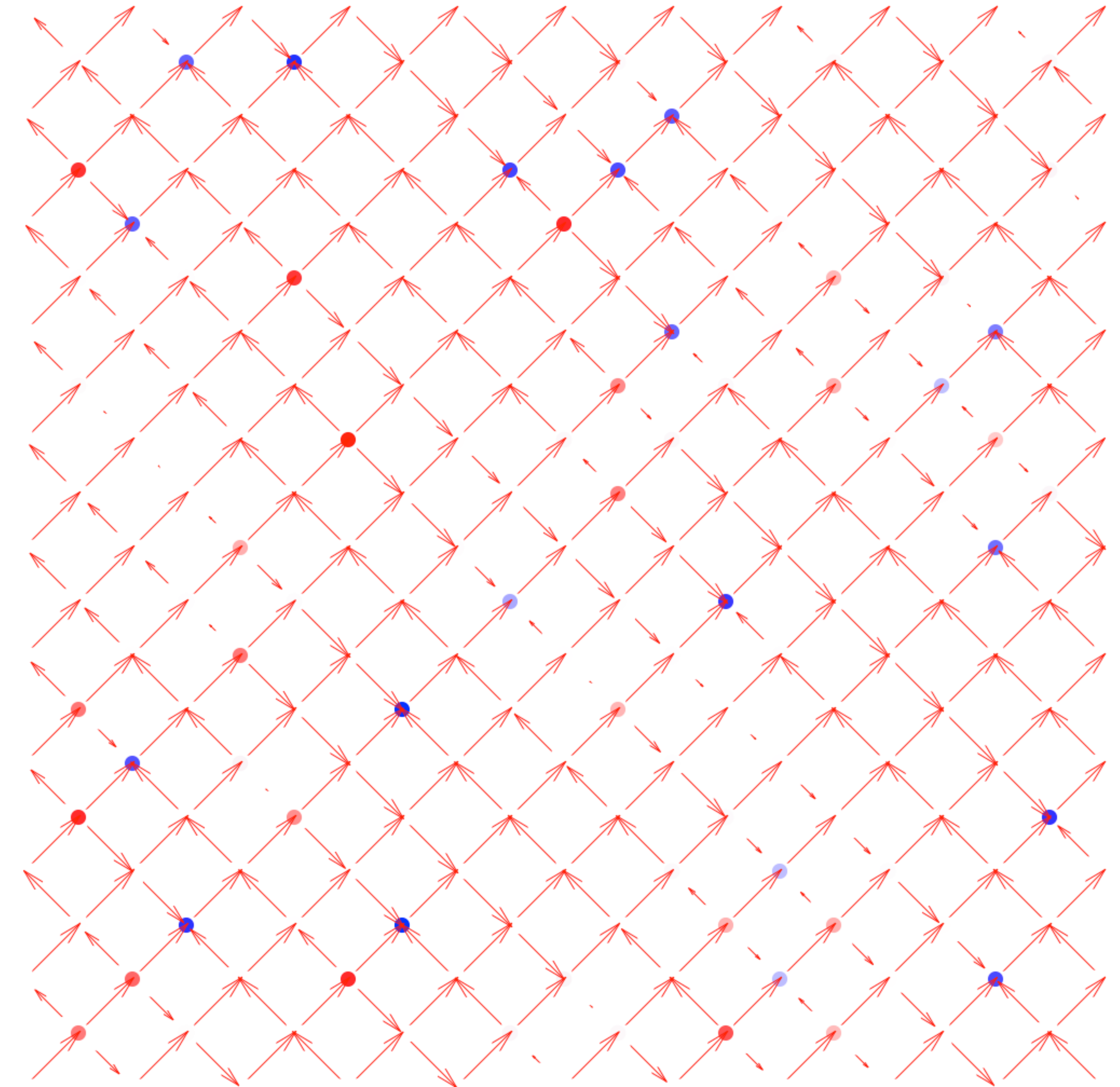}
    \caption{A snapshot of a simulation, for $h=0.48mm$(the degeneracy point) at a fixed external magnetic field. Red and blue represent the negative and positive T3 vertices, e.g. the monopoles. The lengths of the arrows represent the projection of the plane of the simulated rotors.}
    \label{fig:expsim}
\end{figure}
A video of the annealing is available at an online repository for L=20, while a snapshot is shown in Fig. \ref{fig:expsim}.\footnote{Simulation video available at \href{https://www.youtube.com/watch?v=sgVFy4ghFvs}{https://www.youtube.com/watch?v=sgVFy4ghFvs} at the degeneracy point.} The value of $q$ has been chosen to be $1.5$, in order to obtain a crossover of the populations such that for $h\approx 0.55$ the populations of Type I and Type II are identical, which is consistent with experimental results.
\ \\
\begin{figure}[ht!]
    \centering
    \includegraphics[scale=0.2]{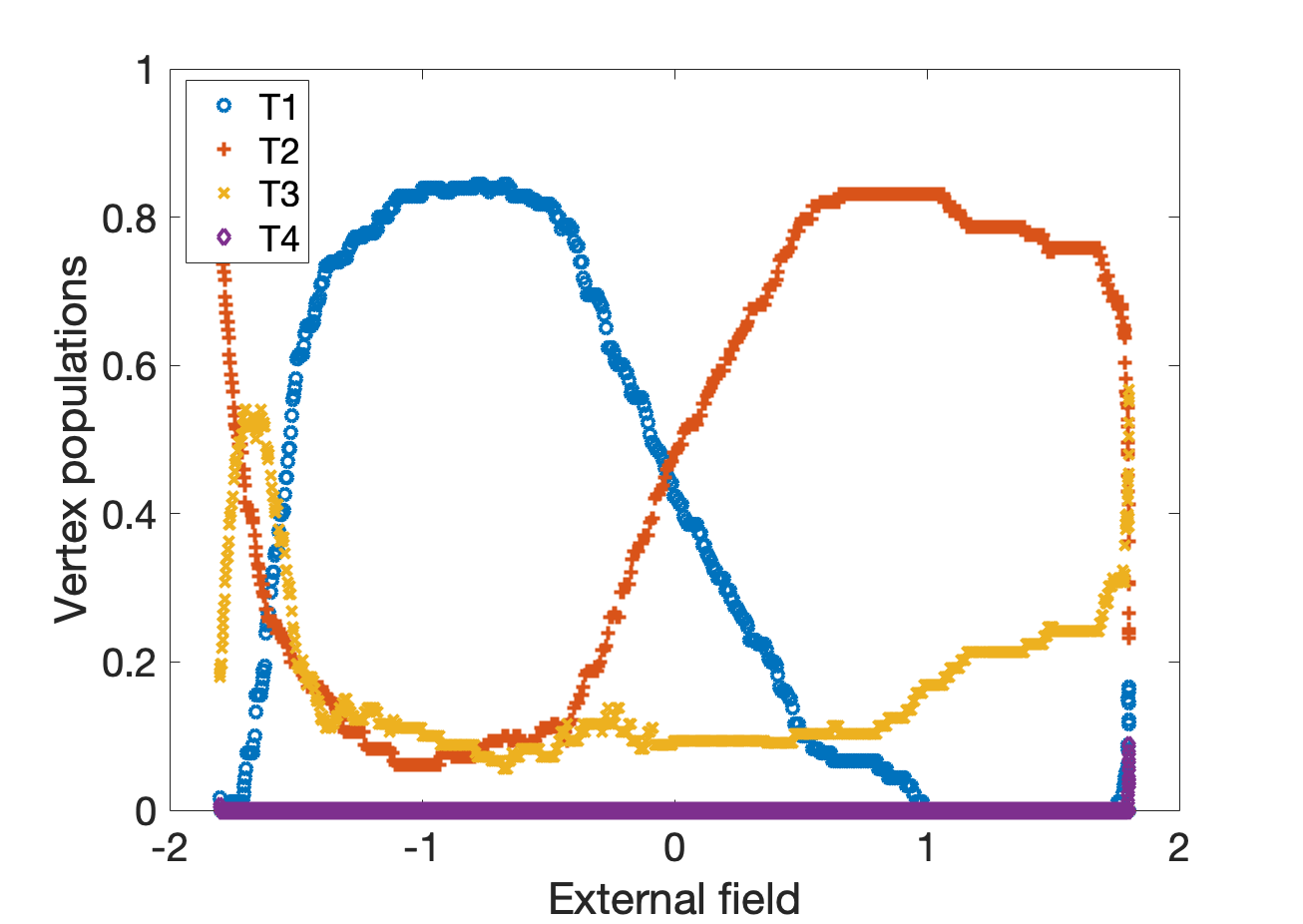}\\
    \includegraphics[scale=0.2]{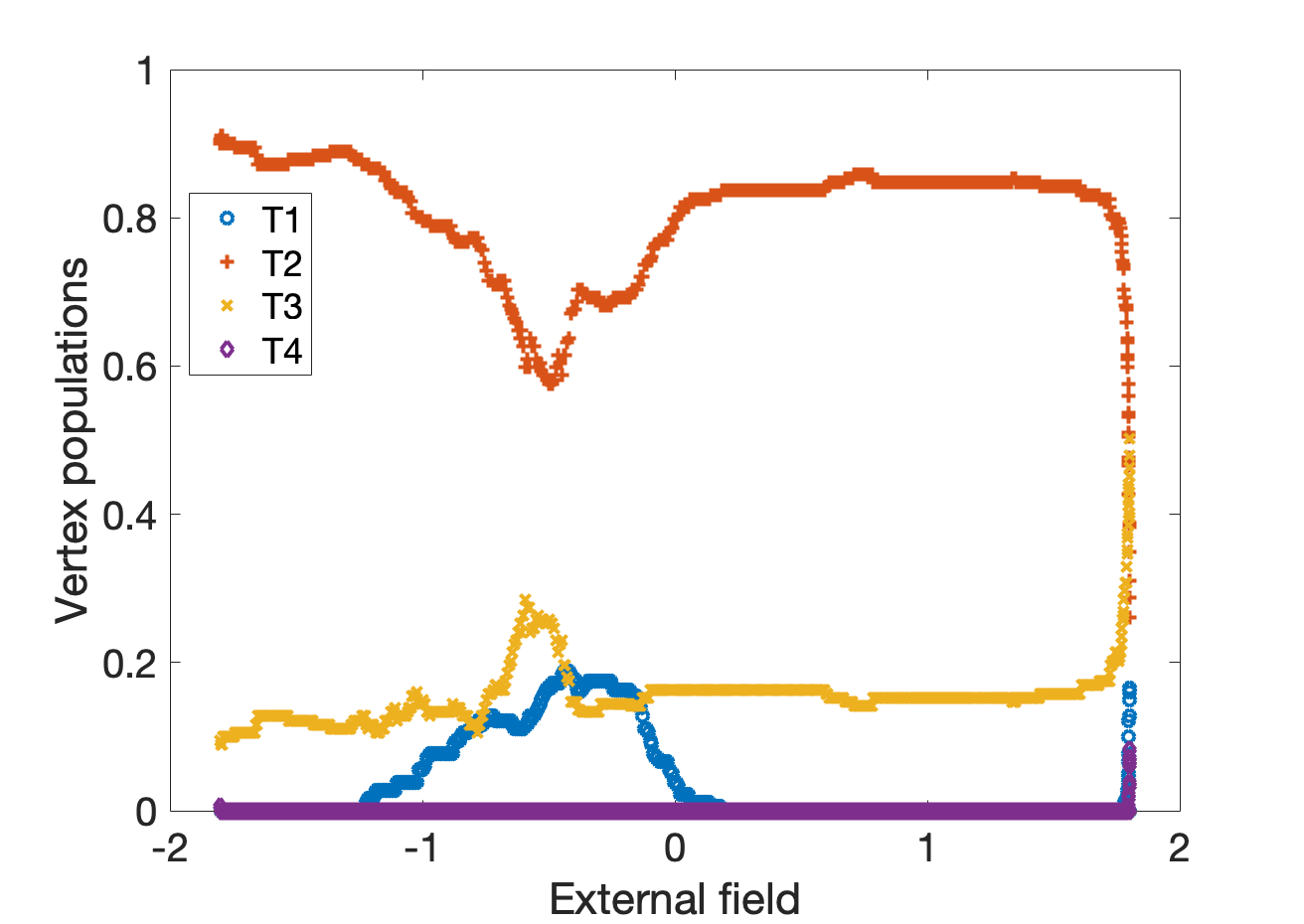}\\
    \caption{Population versus the field strength $m(t)$, $(m_0=1)$ for two values of $h$. On the top we have the populations for $h=0$, where we see that Type I and Type II cross in population as a function of the external field. On the bottom, we plot the populations for $h=0.8$, where we see that Type II are stable as a function of $h$. This is also consistent with the experimental observations.}
    \label{fig:simann}
\end{figure}
Once all the experimental parameters have been found, we can use molecular dynamics to estimate more precisely the location of the degenerate point, as function of $h$, in which the energies of the the Type I and Type II vertices are equal. Similarly to the experiments, we have a crossover between Type I and Type II at around $h=44$ (mm), with the presence of Type III vertices increasing as a function of $h$. During the annealing procedure, the vertices being excited differ depending on the height parameter. This can be seen in Fig. \ref{fig:simann}, where different annealing curves for the vertex populations are observed. For $h=0$, Type I and Type II populations cross as we swipe the external field diagonally across the lattice. For larger values of $h$, however, Type II vertices become more stable, indicating a crossover in energies. These numerical simulations are also consistent with experimental observations. 

\begin{figure}
    \centering
    \includegraphics[scale=0.29]{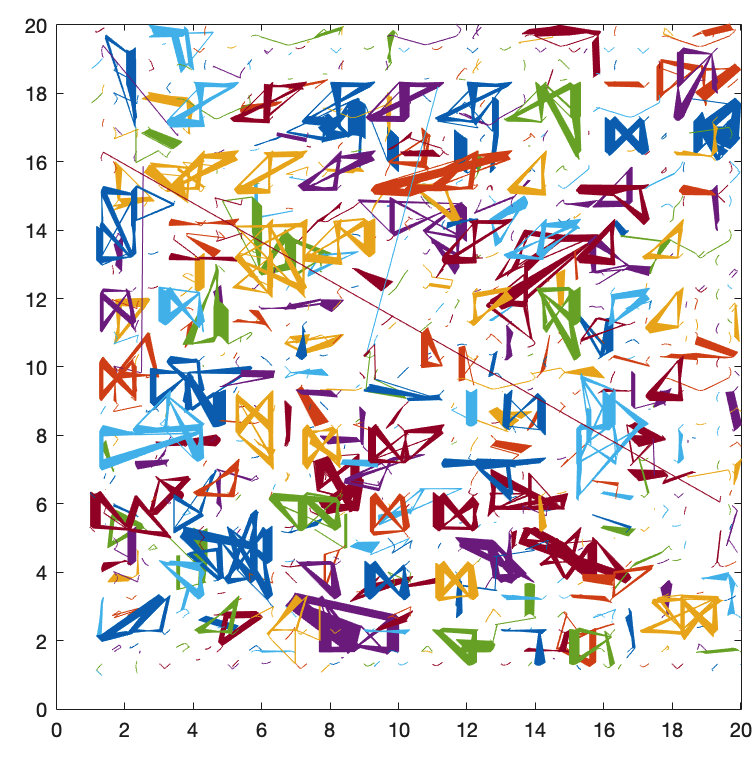}\\
    \includegraphics[scale=0.29]{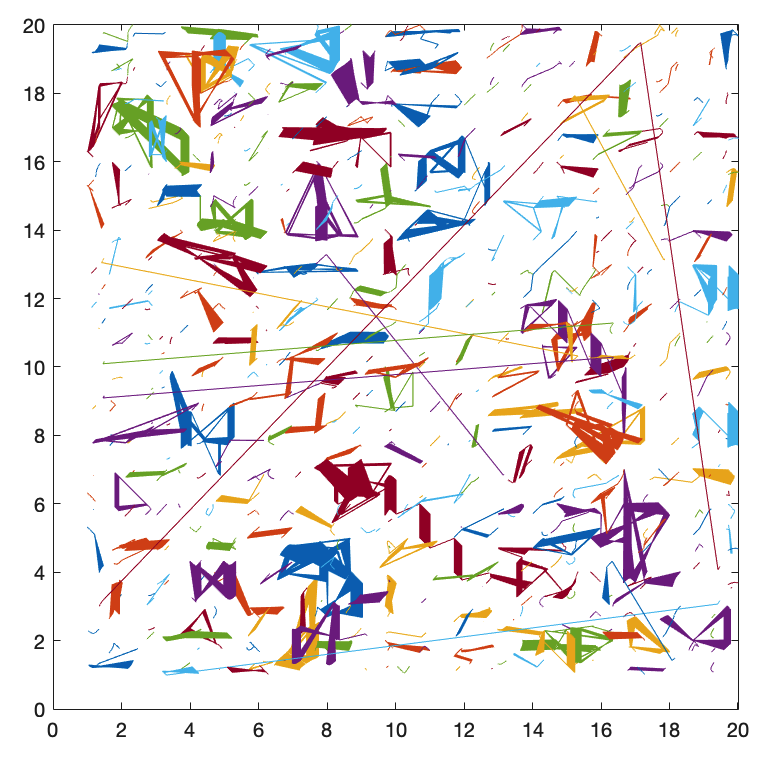}
    \caption{Tracked monopoles in the simulation of a $20\times 20$ lattice, after an annealing procedure. We note that for $h=0$ there is the formation of a higher number of monopoles. For $h=0.5$, there is a lower density of monopoles, but these can move ballistically.}
    \label{fig:monmov}
\end{figure}

We studied numerically the evolution of the systems under a hysteresis loop, comparable to the experiments depicted in Figures 4-6. In this case, the field was applied parallel to one set of spins and perpendicular to the other to match the experimental protocol while the height offset was set to $h = 0.45,$ and 1 respectively to match the experimental vertex type populations. The average in plane magnetization, as plotted in Figures 4-6, is a good fit to the experimental values. A subjective feature relevant to our understanding of these systems is that the degenerate system evolves in leaps at a time when a threshold of external field is cleared. Since this feature is present in both the experiment and simulations, it is relevant to consider as a new dynamical feature of the system's collective motion. 

Finally, we tracked the movements of the Type III vertices, e.g. the monopoles, comparing their movement both at $h=0$ and for $h=0.5$, e.g. near the overlapping points. We tracked every single monopole in a simulation. If no monopoles were created or annihilated, we could simply consider the closest monopole in a successive frame to be the same monopole. However, if the distance between closest monopole was greater than a lattice spacing, the monopole was considered annihilated as no monopole traveled farther in a frame. Finally, the remaining monopoles with no previous time pairs were considered new monopoles. We plotted them over time in Fig. \ref{fig:monmov}. Each colored curve represents the position of a monopole. The first observation is that we can in fact track the monopoles in continuous space rather than discrete: this is due to the fact that as a particular dipole flips, the monopole continuously moves from one site to another. For $h=0$, then, we note that the monopoles are localized, e.g. they are created and destroyed without significant motion from site to site and thus have a finite lifetime. In terms of movement, for $h=0$ we observe only a handful of monopoles moving ballistically. On the other hand, for $h=0.5$, while there is a lower density of monopoles, these can move over longer ballistic trajectories, implying that at this particular point the monopoles are less bound, as we should expect. 

\end{document}